\documentclass{article}
\usepackage[a4paper, margin=1in]{geometry}

\usepackage{blindtext}
\usepackage{hyperref}

\usepackage{makecell}
\usepackage{amsmath}
\usepackage{amsfonts}
\usepackage{bbm}
\usepackage{subcaption}
\usepackage{graphicx}
\usepackage{xcolor}
\usepackage{booktabs}
\usepackage{float}
\usepackage{multirow}
\usepackage[ruled,vlined,linesnumbered]{algorithm2e}
\usepackage{amsmath} 
\usepackage{tikz}
\usepackage{csquotes}
\usepackage[
  backend=biber,
  style=authoryear,
  maxcitenames=2,
  mincitenames=1,
  maxbibnames=6,
  minbibnames=3,
  giveninits=true,
  uniquelist=false
]{biblatex}
\begin{document}
\title{Ultra-Fast Muon Transport via Histogram Sampling on GPUs
}

\author{
\makebox[\textwidth][c]{%
\begin{minipage}{0.9\textwidth}
\centering
Luís Felipe P. Cattelan\textsuperscript{1,2,\textdagger}, 
Shah Rukh Qasim\textsuperscript{1,2}, \\
Patrick H. Owen\textsuperscript{1,2}, 
Nicola Serra\textsuperscript{1,2} \\[4pt]
\footnotesize
\textsuperscript{1}\,Physik-Institut\\
\textsuperscript{2}\,Department of Mathematical Modeling and Machine Learning\\
University of Zurich, Switzerland
\end{minipage}%
}
}
\date{}
\maketitle
\begingroup
  \renewcommand{\thefootnote}{\fnsymbol{footnote}}
  \footnotetext[2]{Corresponding author: lfp.cattelan@cern.ch} 
\endgroup
\begin{abstract}
We present a GPU-accelerated method for muon transport based on histogram sampling that delivers orders of magnitude faster performance than CPU-based Geant4 simulation. Our method employs precomputed histograms of momentum loss and scattering, derived from detailed Geant4 simulations, to statistically reproduce all the non-decaying physics processes during muon traversal through matter. Implemented as a CUDA kernel, the parallel algorithm enables the concurrent simulation of tens of thousands of particles on a single GPU whilst taking into account a complex geometry and a magnetic field force integrated using a fourth-order Runge-Kutta method. Validation against Geant4 in both simple and realistic detector geometries shows that the approach preserves key physical features while achieving speedups of several orders of magnitude, even compared to CPU-based simulations on a large CPU farm with over a thousand cores. This work highlights the significant potential of GPU-based implementations for particle transport, with applicability extending to neutrino propagation and future implementations including discrete processes such as particle decay.
\end{abstract}

\section{Introduction}
\label{sec:introduction}

Monte Carlo (MC) simulations are a cornerstone of computational science, widely used to model complex stochastic processes across domains ranging from medical physics to high-energy particle transport. Their ability to model the dynamics of complex systems from first principles makes them indispensable in diverse applications such as financial modeling~\parencite{boyle1977_options}, chemical kinetics~\parencite{Gillespie1976}, epidemiology~\parencite{Eubank2004}, and particle physics. However, this accuracy comes at a steep computational cost. Since MC methods rely on repeated random sampling of probabilistic processes, achieving statistically significant results often requires simulating millions or even billions of independent events, posing substantial computational challenges for large-scale or iterative tasks such as design optimization or uncertainty quantification.

In particle physics, Geant4~\parencite{Agostinelli2003} has become the de facto standard toolkit for simulating the passage of particles through matter. It is the backbone of nearly all primary simulation frameworks used in major high-energy physics experiments at the European Organization for Nuclear Research (CERN), enabling precise modeling of detector geometries, particle–material interactions, and detector responses. The impact of Geant4 extends far beyond collider physics: it is a key tool in medical-physics applications such as radiotherapy~\parencite{Jiang2004}, dosimetry~\parencite{Jia2025}, and nuclear medicine~\parencite{Freudenberg2011}, as well as in space-science applications, including spacecraft radiation shielding, cosmic-ray modeling, and planetary radiation studies~\parencite{ASAI2008}. Despite its versatility and accuracy, Geant4’s detailed modeling incurs substantial computational overhead, often limiting its use in real-time or large-parameter-space studies.

To overcome this limitation, two complementary strategies have emerged. The first explores machine-learning–based surrogate models that emulate the statistical behavior of complex simulations. Notably, \textcite{Paganini2018} demonstrated the use of Generative Adversarial Networks (GANs) to reproduce calorimeter shower images with orders-of-magnitude faster inference. Following this, several works have extended this approach to diverse detector geometries and particle types~\parencite{vallecorsa2018generative, carminati2018three}, and investigated their integration into full experimental simulation chains~\parencite{atlas2024deep, ratnikov2020generative}. Complementing these developments, recent works have discussed the potential and limitations of generative models in particle physics~\parencite{das2024understand, adelmann2022new}. The second strategy focuses on hardware acceleration---redesigning or reimplementing the core transport algorithms to exploit massively parallel architectures such as Graphics Processing Units (GPUs). GPUs, with their thousands of lightweight cores, are well-suited to the inherently data-parallel nature of MC simulations, where each particle’s trajectory can be computed independently.

Early GPU-based efforts developed specialized standalone MC codes for medical and radiation-transport applications, including MC-GPU~\parencite{badal2009accelerating} for photon transport, and gPMC and G4CU~\parencite{jia2012gpu,murakami2013geant4} for dose calculations in proton therapy, achieving speedups of up to two orders of magnitude relative to traditional CPU implementations. More recently, hybrid frameworks have aimed to bring GPU acceleration to general-purpose particle-transport systems. The AdePT framework~\parencite{amadio2023offloading} integrates with Geant4 to offload electromagnetic shower propagation to GPUs, while the Celeritas project~\parencite{tognini2022celeritas,Lund2025} takes a ground-up GPU-first approach, achieving more than an order-of-magnitude improvement over CPU-based Geant4 transport while maintaining physics consistency within statistical uncertainty. Together, these efforts demonstrate the growing maturity and promise of GPU-accelerated Monte Carlo transport.

One particularly demanding use case that motivates our work is the optimization of active muon shielding systems in beam-dump experiments, such as the SHiP Active Muon Shield~\parencite{muon_shield_2017}. Designing such systems requires simulating the transport of billions of muons through complex geometries of magnets and absorbers to minimize background in the downstream detector. Full Geant4 simulations are prohibitively expensive in this context: even simplified studies have required several weeks of computation on large CPU clusters, limiting the scope of design-space exploration and increasing the risk of suboptimal configurations. These challenges underline the need for faster, physically faithful transport simulations.

In this work, we present a GPU-accelerated Monte Carlo particle-transport framework based on histogram sampling. Our method leverages precomputed histograms of momentum loss and scattering, derived from detailed Geant4 simulations, to reproduce the statistical behavior of particle–matter interactions with high fidelity. Implemented as a CUDA kernel, the algorithm simulates tens of thousands of particles in parallel, achieving speedups of several orders of magnitude compared to CPU-based Geant4, while preserving key physical features.

\begin{figure}
    \centering
    \begin{subfigure}[b]{0.44\textwidth}
        \centering
        \raisebox{0.8cm}[0pt][0pt]{\includegraphics[width=\linewidth]{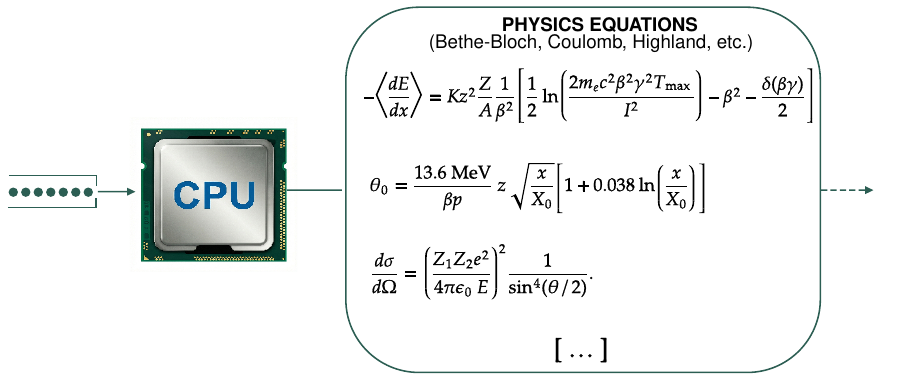}}
    \end{subfigure}
    \hfill
    \begin{subfigure}[b]{0.55\textwidth}
        \centering
        \includegraphics[width=\linewidth]{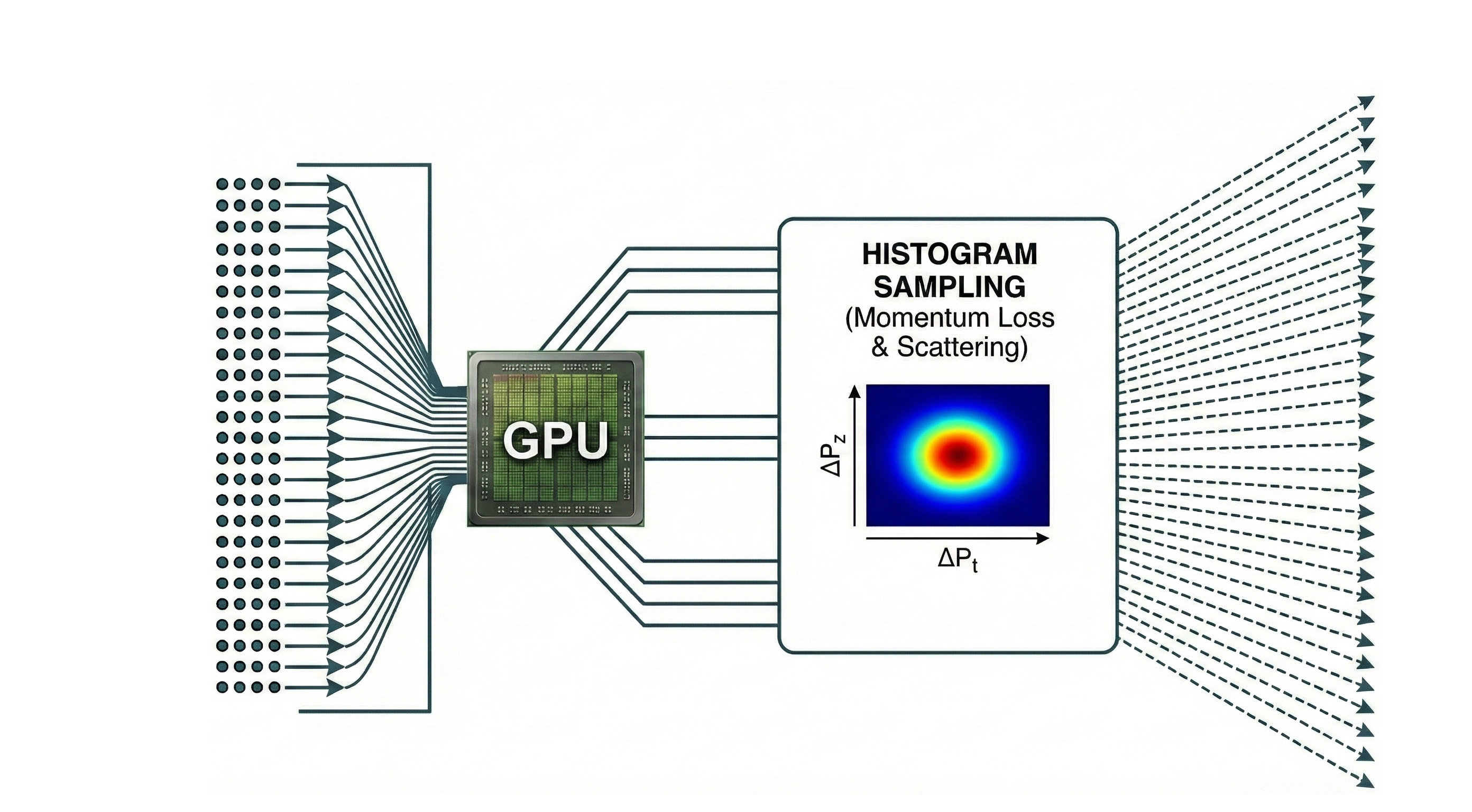}
    \end{subfigure}
    \caption{Schematic comparison between traditional Geant4 sequential transport (left) and our GPU-parallelized histogram sampling method (right).}
    \label{fig:methods_schematic}
\end{figure}

The remainder of this work is structured as follows. The Methodology section introduces the method framework and the key concepts. Section~\ref{sec:building_histograms} details the construction of histograms of energy loss and scattering and the momentum-binning strategy. Section~\ref{sec:simulation} presents the GPU-based transport algorithm and implementation details. The Validation and benchmarking section compares results against Geant4 in both simplified and realistic detector geometries and reports timing studies. The Discussion and Conclusion section summarizes implications and outlines opportunities for future extensions.

\section{Methodology}

Our approach to accelerating particle transport simulations is based on a data-driven heuristic implemented for massively parallel execution on GPUs. The method requires defining key simulation parameters upfront: the specific set of materials involved, the desired particle momentum range, and a fixed step distance $d$\footnote{The method can be adapted to consider a dynamic step size, as done in Geant4.}. This step distance acts as a crucial hyperparameter, mediating a trade-off between simulation speed (larger $d$) and physical precision (smaller $d$). 

Crucially, the simulation model presented in this work focuses solely on the effects of momentum loss and scattering due to interactions within materials (captured by the histograms) and the influence of external forces (specifically, the electromagnetic Lorentz force). Other physical processes inherent to a full Monte Carlo simulation, such as particle decay, are not considered in this implementation.

The overall process consists of two main phases: an offline pre-computation phase using these predefined parameters and an online GPU-accelerated simulation phase. The implementation code and experiments details can be found in \url{https://github.com/lfpc/cuda_muons}.

\subsection*{Offline Phase} This initial phase leverages the high-fidelity physics models of the Geant4 simulation toolkit to generate training data based on the chosen materials, momentum range, and step size $d$. For each relevant material and a series of discrete momentum bins, we simulate a large number of particles traversing the fixed distance $d$, recording the resulting longitudinal momentum change and transverse momentum kick. This data captures the complex, stochastic effects of energy loss and multiple scattering. 

The collected data is then processed to build two-dimensional histograms for each material and momentum bin. These histograms, when normalized, represent the joint discrete probability distribution of the momentum changes.

\subsection*{The Alias Method} Since sampling from these histograms is the most frequent operation during the simulation, a naive search with $O(\log n)$ complexity would introduce a significant bottleneck. To resolve this, we employ the Alias Method~\parencite{alias1977walker}. This technique transforms the uneven histogram probabilities into a structure of $N$ equiprobable bins, each containing at most two outcomes (a primary and an alias). We pre-compute the associated lookup tables ($\mathtt{Prob}$ and $\mathtt{Alias}$), enabling the GPU threads to draw samples in constant time $O(1)$ during the simulation.

\subsection*{Online GPU-Accelerated Simulation Phase} The core simulation is implemented as a CUDA (Compute Unified Device Architecture) kernel, running entirely on the GPU. Unlike CPUs, which optimize for low latency using a few powerful cores, GPUs utilize thousands of efficient cores to maximize throughput.  This architecture is ideal for Monte Carlo simulations, which are inherently data-parallel: the history of one particle is independent of another.

We leverage this by assigning each particle to a single GPU thread, allowing tens of thousands of particles to be simulated concurrently. Each thread executes the full transport logic for its assigned particle through the following iterative process:

\begin{enumerate}
    \item Determine the material at the particle's current position using geometry lookup functions.
    \item For the identified material, determine the correct momentum bin and apply the momentum changes by sampling from the corresponding pre-computed histogram using the alias method.
    \item Calculate the effect of external electromagnetic fields using a fast grid lookup and integrate the particle's equation of motion to update its position and momentum over the fixed step length $d$.
\end{enumerate}
This process repeats until the particle meets a termination condition (e.g., falls below a minimum momentum threshold or reaches a specific region). The simulation algorithm is detailed in Section \ref{sec:simulation}.

\section{Building histograms}
\label{sec:building_histograms}

To efficiently simulate the transport of particles through matter, we precompute histograms that capture the statistical behavior of energy loss and scattering for different materials and momentum ranges. The division of the histograms into momentum bins allows us to account for the energy-dependent nature of particle interactions, ensuring that our simulations remain accurate across a wide range of particle energies. More details on how to divide these momentum bins can be found in section \ref{sec:momentum_binning}.

For a given material $\mathcal{M}$, a particle (in this work we only consider muons), and a given momentum bin $i$, we simulate N particles with initial momentum $\mathbf{P} = [0,0,P_z],\quad P_z \sim \mathcal{U}(p_i, p_{i+1})$ traveling through this material for a distance $d$ and record their momentum loss $\Delta P_z$ and the scattering through the transverse momentum $\Delta P_t = \sqrt{P_x^2 + P_y^2}$ (note the initial transverse momentum of the particle is always set as 0). Differently than Geant4, we consider a fixed step size $d$, instead of sampling the distance to the next interaction. This is a simplification that allows us to build the histograms more easily, but it is not a limitation of our method, as one can build histograms for different step sizes if needed.

\begin{figure}[ht]
    \centering
    \includegraphics[width=0.9\textwidth]{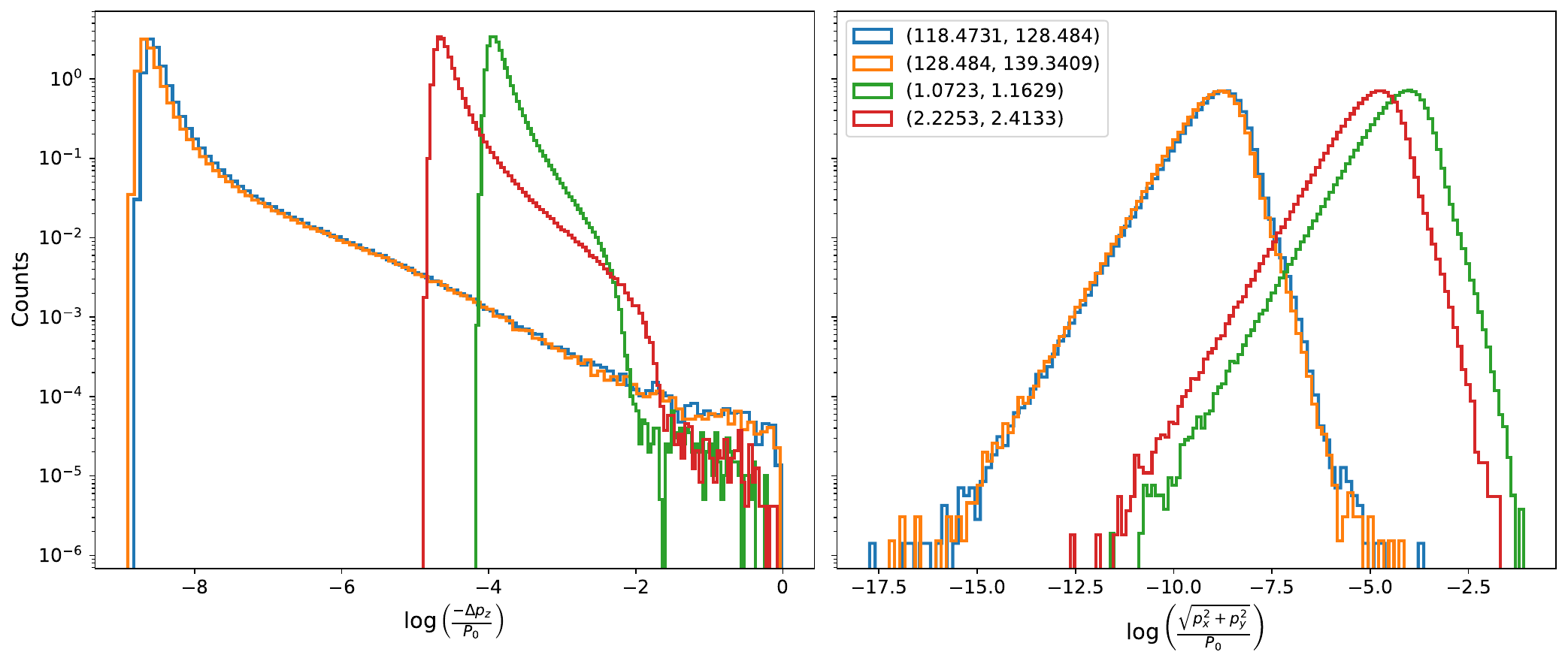}
    \caption{Histograms of momentum loss (left) and scattering (right) for different momentum bins.}
    \label{fig:histograms_1d}
\end{figure}

To ensure that our sampling is physically realistic, we first normalize the momentum loss and scattering values, binning them as a fraction of the particle's initial momentum. This simple step prevents the simulation from ever sampling a momentum loss greater than what the particle actually has.
More importantly, we used a logarithmic scale for these bins. This is because the underlying physics of particle interactions naturally span several orders of magnitude, from tiny energy transfers to catastrophic ones, as shown in Figure \ref{fig:histograms_1d}.

\begin{figure}[h]
    \centering
    \includegraphics[width=0.7\textwidth]{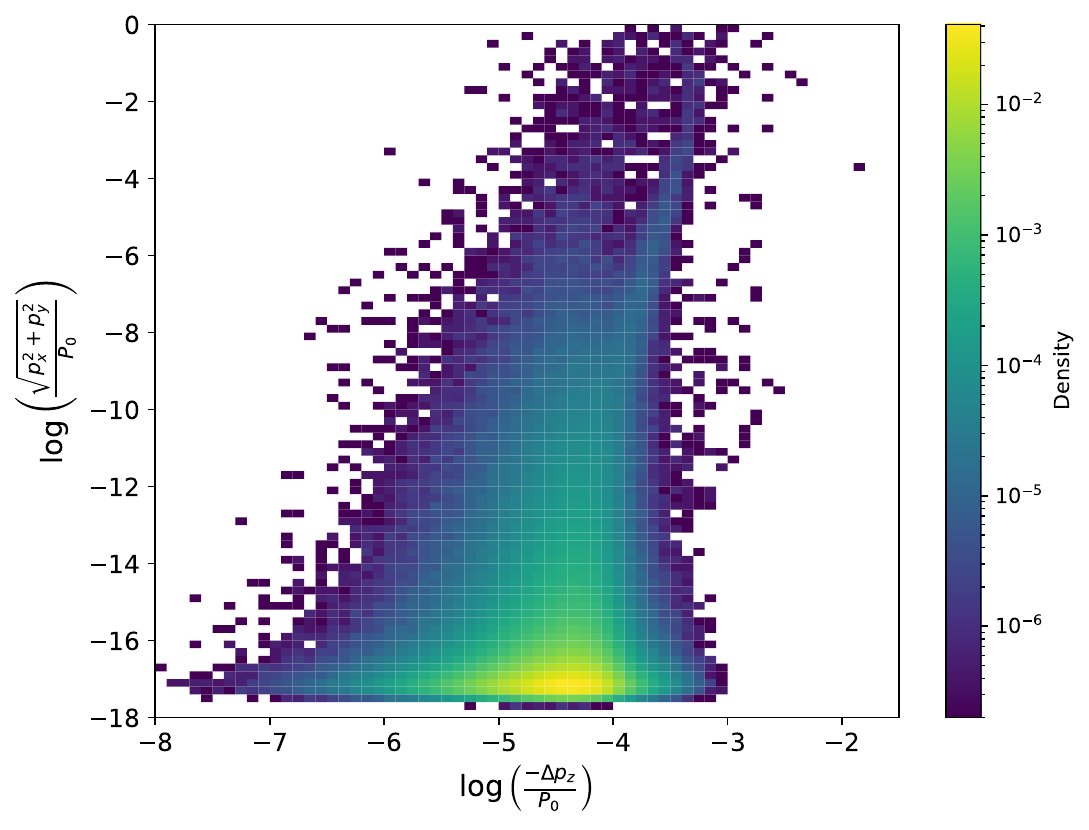}
    \caption{2D histogram of momentum loss vs scattering for muons with momentum between 118.47 and 128.48 GeV traveling through iron for a distance of 2 cm.}
    \label{fig:histogram_2d}
\end{figure}

From the data collected, we build a 2D histogram $H_{\mathcal{M},i}$ relating the momentum loss, $\log\left(- \frac{\Delta P_z}{|\mathbf{P}|}\right)$ and the scattering  $\log\left(\frac{\Delta P_t}{|\mathbf{P}|}\right)$. The histogram is then normalized to obtain a discrete probability distribution function and, finally, an alias table is built for efficient sampling during the particle transport simulation. Figure \ref{fig:histogram_2d} shows an example of the 2D histogram built for muons with momentum between 118.48 and 128.48 GeV traveling through iron for a distance of 2 cm.

For all the analysis presented in this work, we used $N=5\times10^6$ particles to build each histogram, since this was enough to obtain a smooth distribution within a reasonable computation time. The generation of all histograms for a given material can be done in a couple of hours on a single CPU core, and it is a one-time cost that can be reused for multiple simulations.

\subsection{Momentum binning}
\label{sec:momentum_binning}

A critical aspect of our method is the division of the momentum spectrum into discrete bins. Figure \ref{fig:histograms_1d} shows the histograms for four distinct momentum bins. The figure highlights how interaction outcomes can be highly sensitive to the particle's initial momentum in some regions while showing little variation in others. Hence, a uniform binning strategy would either miss important variations in low-momentum regions or waste computational resources in high-momentum regions where changes are minimal. For that reason, we employ a non-uniform binning strategy that allocates more bins to momentum ranges where interactions are more sensitive and fewer bins where they are less so. To better understand how to choose these bins, we need to understand how the energy loss varies with momentum.

 The mean rate of energy loss (mean stopping power), more precisely denoted as $\langle dE/dx \rangle$, is described mainly by the Bethe-Bloch formula:
\begin{equation}
    -\left\langle\frac{dE}{dx}\right\rangle = K z^2 \frac{Z}{A} \frac{1}{\beta^2} \left[ \frac{1}{2} \ln \frac{2m_e c^2 \beta^2 \gamma^2 W_{max}}{I^2} - \beta^2 - \frac{\delta(\beta\gamma)}{2} \right]
\end{equation}

At low momenta, the $1/\beta^2$ term causes the mean energy loss to be very high and change rapidly. As momentum increases, the loss reaches a minimum before starting a slow, logarithmic rise. This means the average behavior is highly sensitive in some regions and more stable in others, explaining the behavior observed in Figure~\ref{fig:histograms_1d}. 

One possibility to create the dynamic binning is to use the derivative of the Bethe-Bloch formula to identify regions where the energy loss changes rapidly. Alternatively, one could perform this analysis empirically by running preliminary simulations to observe where significant variations occur. These approaches are discussed in Appendix \ref{appendix:momentum_binning}. In this work, we chose a simpler approach by using a logarithmic binning strategy that allocates more bins to the low-momentum region where the energy loss changes rapidly. Besides the simplicity of implementation, using a logarithmic scale allows for faster indexing during the simulation, as it can be computed using basic arithmetic operations, instead of requiring a search through an array of bin edges.

\begin{figure}[h]
    \centering
    \begin{subfigure}[b]{0.95\textwidth}
        \centering
        \includegraphics[width=\textwidth]{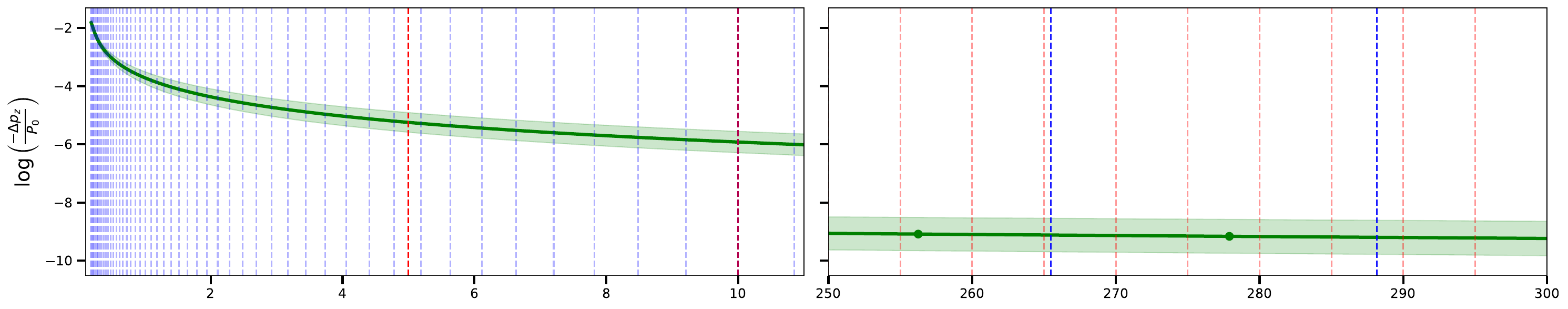}
    \end{subfigure}
    \begin{subfigure}[b]{0.95\textwidth}
        \centering
        \includegraphics[width=\textwidth]{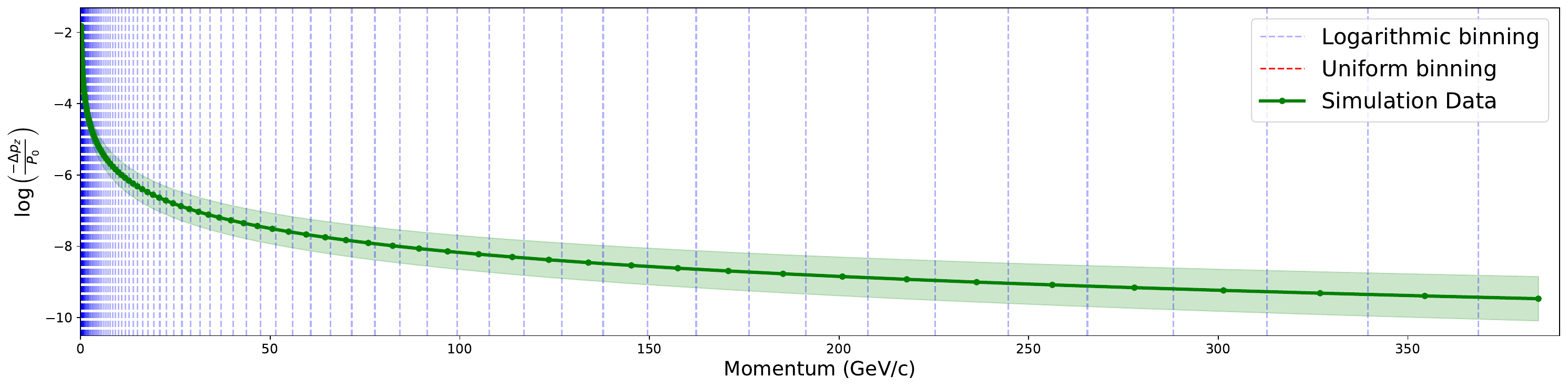}
    \end{subfigure}
    \caption{Plot of average energy loss for muons in iron as a function of momentum, along with two different binning strategies: uniform and logarithmic. Shaded region indicates the standard deviation of the sampled energy loss. Top plots show a zoomed-in view in different momentum ranges.}
    \label{fig:binning}
\end{figure}

In Figure \ref{fig:binning}, we show the sampled energy loss for muons in iron as a function of momentum. Alongside that, we plot two different binning strategies: a uniform binning, with bin size of 5 GeV, and a logarithmic binning that allocates more bins to the low-momentum region where the energy loss changes rapidly. As can be seen, the logarithmic binning captures the variations in energy loss more effectively across the entire momentum range. Hence, it is the preferred choice for our histogram construction.

\section{Simulation}
\label{sec:simulation}

After building the histograms for each material and momentum bin, we can proceed to simulate the transport of particles through a given geometry. The simulation consists of tracking each particle's trajectory step-by-step, updating its momentum and position based on the sampled energy loss and scattering from the precomputed histograms, as well as the influence of any electromagnetic fields present in the environment.

\begin{enumerate}
    \item \textbf{Material lookup} \\ At each step, the algorithm first determines the material $\mathcal{M}$ at the particle's current position. The specific algorithm for material lookup is dependent on how the detector geometry is represented. For the example where the geometry is defined by a series of convex polyhedra, we use a multi-stage containment test.
    This is achieved by defining a 2D quadrilateral slice of the candidate block at the particle's $z$-height via interpolation, followed by a point-in-polygon test. %
For each edge vector $\vec{e}_i$, we evaluate the 2D cross product component:
\begin{equation}
    S_i = (\vec{e}_i \times (\vec{p} - \vec{v}_i))_z = e_{ix}(p_y - v_{iy}) - e_{iy}(p_x - v_{ix})
\end{equation}
The particle is inside the volume if and only if the signs of all $S_i$ are consistent. To minimize computational cost, candidate geometries are pruned using a spatial hashing grid prior to this detailed check.

    \item \textbf{Energy loss and scattering sampling} \\ Given the identified material and momentum bin $i$, the momentum loss and scattering are sampled from the corresponding histogram $H_{\mathcal{M},i}$ using the alias method. A small random perturbation (jitter) is added to the sampled values to suppress binning artifacts.
    The angle of scattering is assumed to be isotropic, and it is sampled uniformly from $0$ to $2\pi$.
    Since the sampled momentum change, which we denote as $\Delta\vec{p}$, is defined relative to the particle's instantaneous direction of travel, it must be rotated into the global coordinate system. The momentum update is performed using the Rodrigues' rotation formula:
\begin{equation}
    \label{eq:rodrigues}
    \vec{p} \leftarrow \vec{p} + \Delta\vec{p} \cdot \cos\theta +  (\mathbf{k} \times \Delta\vec{p})\cdot\sin\theta + (1-\cos\theta)(\mathbf{k} \cdot \Delta\vec{p})\mathbf{k}
\end{equation}
Here, $\mathbf{k}$ is the unit rotation axis perpendicular to both the reference $z$-axis and the particle's direction $\vec{p}$, and $\theta$ is the angle between them.

\item \textbf{Electromagnetic force} \\ 
The magnetic field $\vec{B}$ is pre-computed using dedicated simulation software and stored on a regular 3D grid, enabling efficient retrieval via nearest-neighbor interpolation.
The particle's trajectory through the detector's electromagnetic fields is governed by the Lorentz force:
\begin{equation}
\frac{d\vec{P}}{dt} = q(\vec{E} + \vec{v} \times \vec{B})
\end{equation}
We solve this equation of motion numerically using a fourth-order Runge-Kutta (RK4) integrator to update the particle's state.
\end{enumerate}

Taking all of these components into account, we can now outline the complete simulation algorithm. For a given particle, the simulation algorithm proceeds as follows:

\begin{algorithm}[H]
\DontPrintSemicolon
\caption{Particle Transport Simulation Loop}
\label{alg:simulation_loop}
\KwIn{
    Initial position $\mathbf{r}$, initial momentum $\mathbf{p}$\;
    Histograms $\mathcal{H}$, B-field map $\mathbf{B}_{map}$, geometry $\mathcal{G}$\;
    Number of steps $n_{steps}$, step length $d_{step}$, break conditions ($p_{min}$, $\mathbf{r}_{max}$)\;
}
\BlankLine
\For{step = 1 \KwTo n\_steps}{
    $p_{mag} \leftarrow |\mathbf{p}|$\;
    \If{$p_{mag} < p_{min}$ \textbf{or} $z \geq z_{max}$ \textbf{or} $x \geq x_{max}$ \textbf{or} $y \geq y_{max}$}{
        \textbf{break}\;
    }

    $\mathcal{M} \leftarrow$ GetMaterial($\mathbf{r}$, $\mathcal{G}$)\;
    \If{$\mathcal{M}$ is not \text{vacuum}}{
        $i_{bin} \leftarrow$ GetMomentumBin($p_{mag}$)\;
        $\delta_z, \delta_t$ $\leftarrow$ AliasSample($\mathcal{H}_{\mathcal{M}, i_{bin}})+ \text{UniformJitter}(\mathcal{H}_{\mathcal{M}, i_{bin}})$\;
        $\phi \leftarrow \text{RandomUniform}(0, 2\pi)$\;
        $\Delta\mathbf{p} \leftarrow [\delta_t \cdot \cos\phi, \delta_t \cdot \sin\phi, \delta_z]$\;
        $\mathbf{p} \leftarrow \text{Rotate}(\mathbf{p}, \Delta\mathbf{p})$ \tcp*{using Eq. (\ref{eq:rodrigues})}  
    }
    $\vec{B} \leftarrow$ GetField($\mathbf{r}, \mathbf{B}_{map}$)\;
    ($\mathbf{r}, \mathbf{p}$) $\leftarrow$ RK4Step($\mathbf{r}, \mathbf{p}$, $\vec{B}$, $d_{step}$)\;
}
\end{algorithm}

\

The workflow outlined in Algorithm \ref{alg:simulation_loop} is implemented as a standalone CUDA kernel, enabling the massively parallel simulation of particle batches. In this model, each GPU thread is assigned a unique particle and executes its transport loop independently. 
Essential static data—including geometry definitions, magnetic field maps, and pre-computed physics histograms—are pre-loaded into the GPU's global memory to maximize throughput. 

\section{Validation and benchmarking}

We validate the accuracy of our proposed method against those obtained from the Geant4 software and benchmark the time performance. We implemented our method with the fixed step size of 2 cm, which is a reasonable compromise between accuracy and performance, and constructed the momentum bins using the logarithmic binning strategy described in section \ref{sec:momentum_binning}, with base 10 and with 95 bins between 0.18 and 400 GeV. 

\begin{figure}[h]
    \centering
    \begin{subfigure}[b]{0.8\textwidth}
        \centering
        \includegraphics[width=\textwidth]{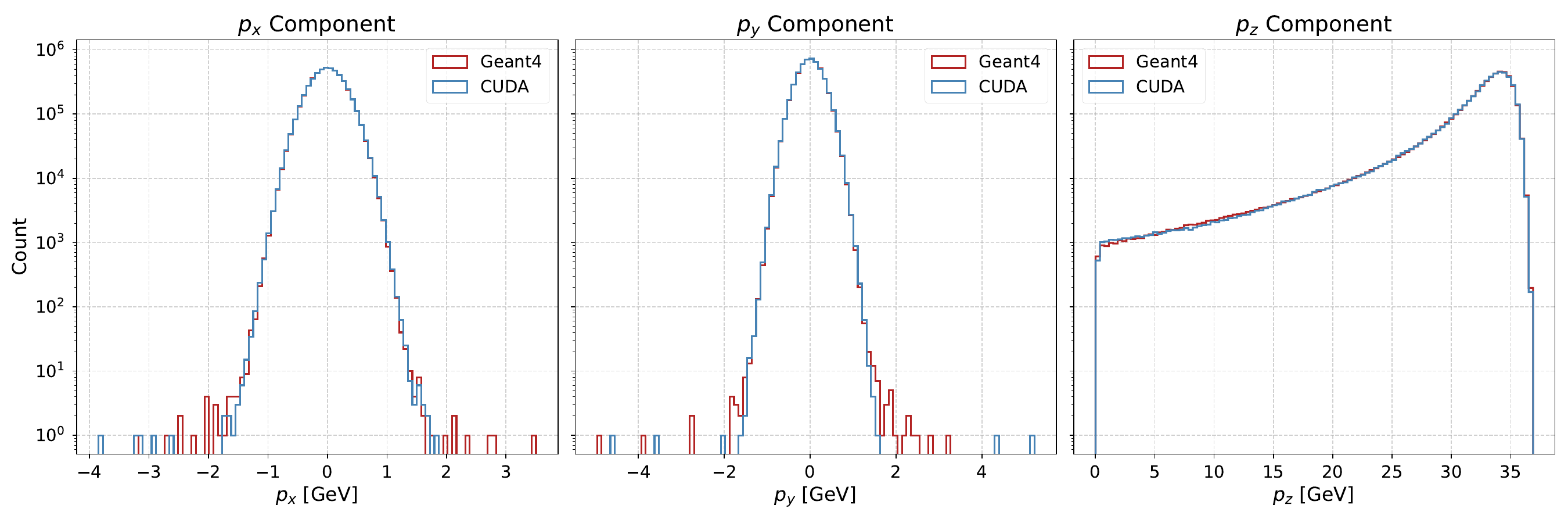}
    \end{subfigure}
    \begin{subfigure}[b]{0.8\textwidth}
        \centering
        \includegraphics[width=\textwidth]{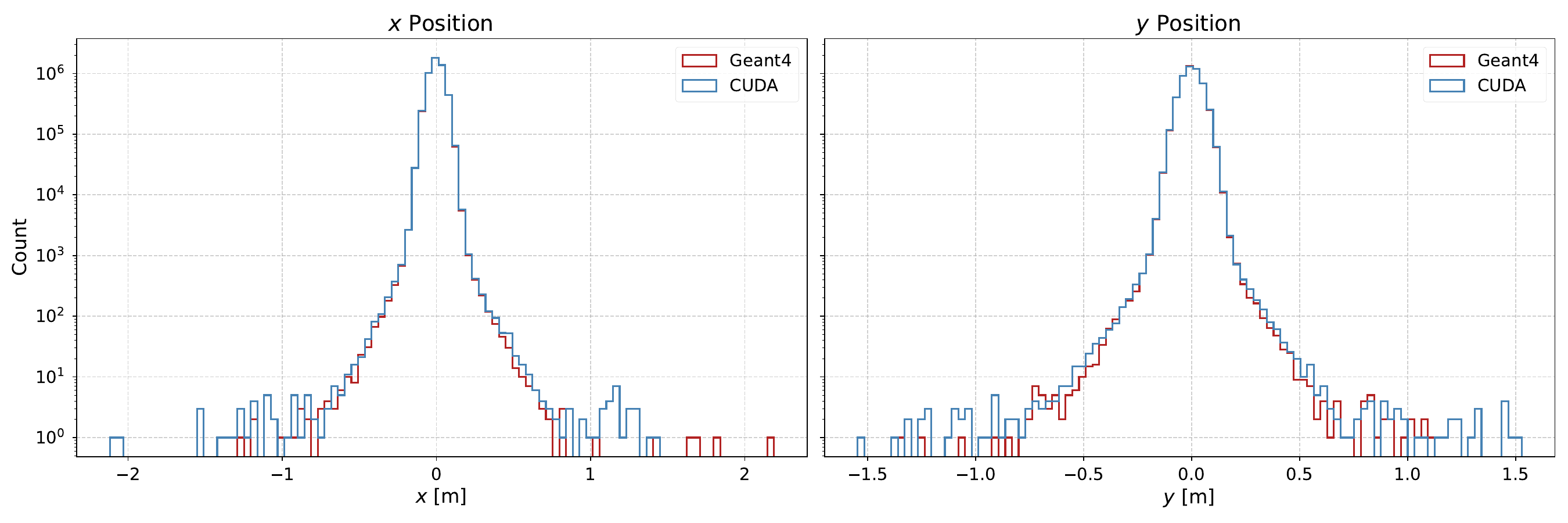}
    \end{subfigure}
    \caption{Validation of our method against Geant4 for muons for a transport distance of 10m.}
    \label{fig:comparison_cuda_geant4}
\end{figure}

We start with a simple scenario where muons with a momentum of 50 GeV travel through an iron block for 10 meters. We simulate the transport of 5 million muons using both our method and Geant4 and compare the final state (i.e., position and momentum) of the particles. The results are shown in Figure \ref{fig:comparison_cuda_geant4}, demonstrating good agreement between the two methods.

\begin{figure}[h]
    \centering
    \includegraphics[width=0.8\textwidth, trim={0 7cm 0 7cm}, clip]{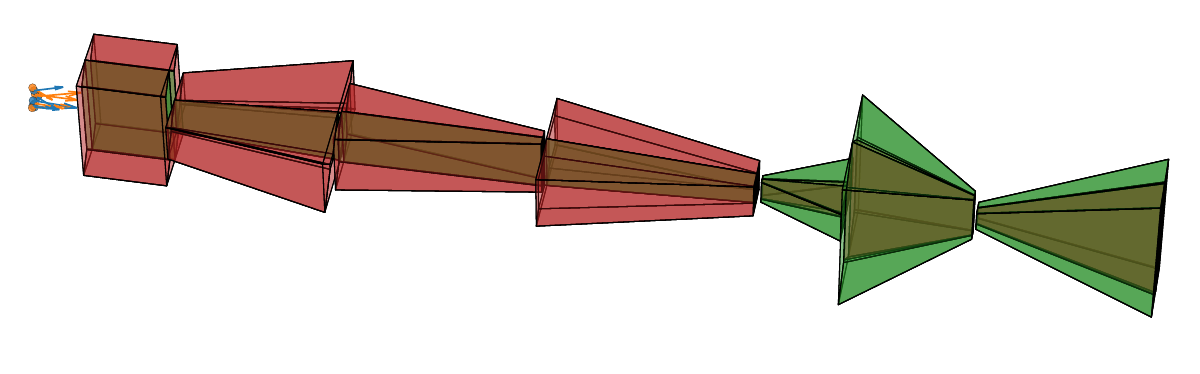}
    \caption{SHiP's Active Muon Shield magnet configuration. Colors indicate the polarization of the magnets. }
    \label{fig:ship_magnets}
\end{figure}

\begin{figure}[ht!]
    \centering
    \begin{subfigure}[b]{0.8\textwidth}
        \centering
        \includegraphics[width=\textwidth]{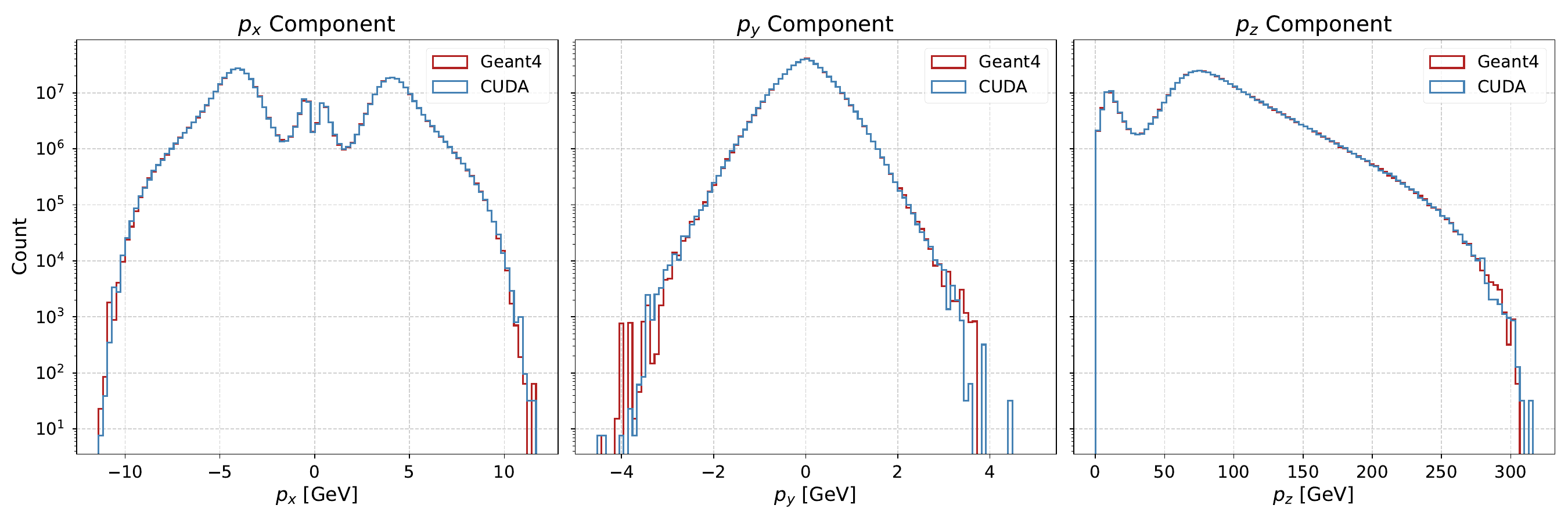}
    \end{subfigure}
    \begin{subfigure}[b]{0.8\textwidth}
        \centering
        \includegraphics[width=\textwidth]{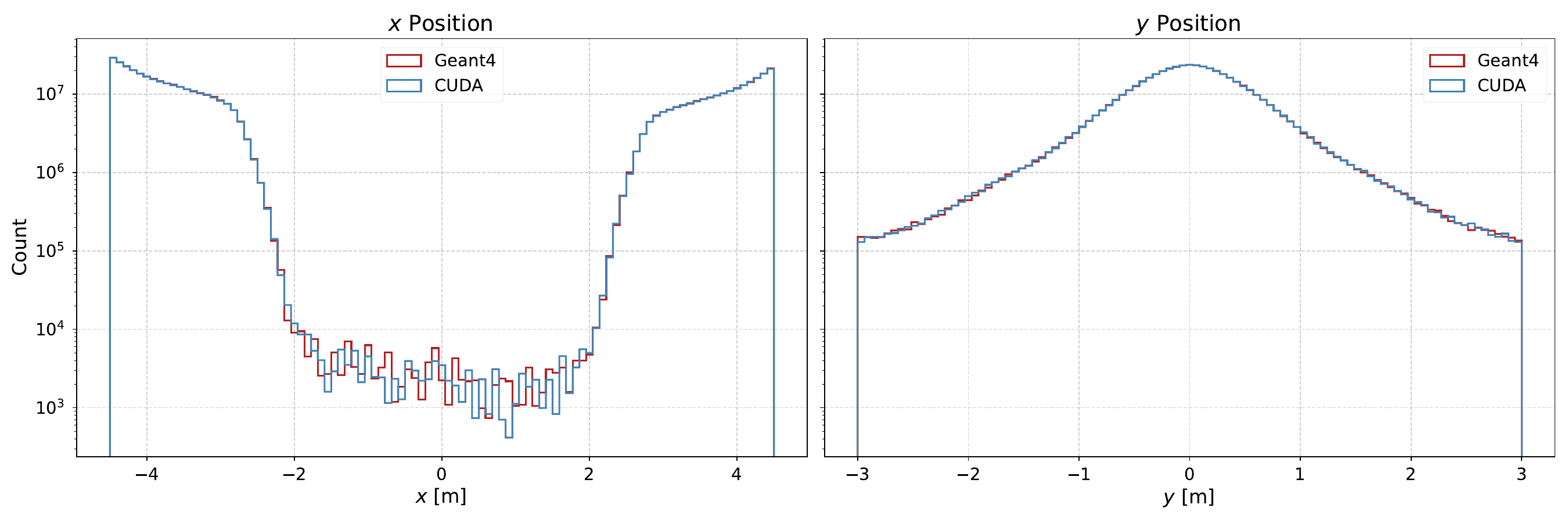}
    \end{subfigure}
    \caption{Validation of our method against Geant4 for the simulation of $5\times10^8$ muons through the SHiP's Active Muon Shield.}
    \label{fig:comparison_muonshield}
\end{figure}

Next, we present a more complex scenario that better represents a realistic use case. Motivated by SHiP's Active Muon Shield~\parencite{muon_shield_2017}, we constructed a sequence of magnets composed of iron blocks, as shown in Figure~\ref{fig:ship_magnets}. Around the magnets, walls made of concrete are also considered. The magnetic field map is pre-calculated using the Snoopy software~\parencite{snoopy} via finite element analysis. We simulate the transport of muons and anti-muons with a broad spectrum of initial momenta through this geometry, and compare the particle distributions at the sensitive plane located at $z=82$\,m relative to the start of the first magnet. As shown in Figure~\ref{fig:comparison_muonshield}, our method reproduces the Geant4 distributions with high fidelity, accurately tracking particles through the alternating magnetic fields and heterogeneous materials.

\begin{figure}[h!]
    \centering
    \begin{subfigure}[b]{0.58\textwidth}
        \centering
        \includegraphics[width=\textwidth]{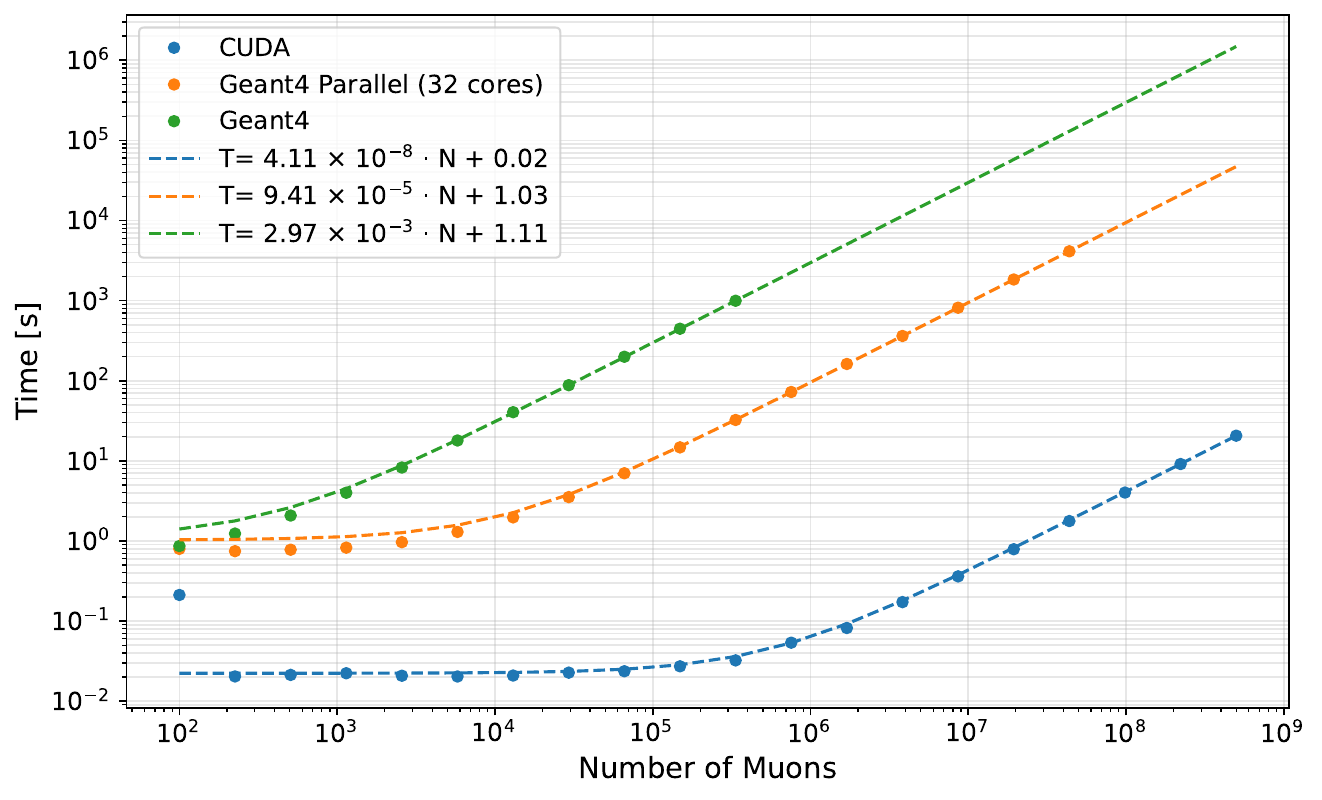}
        \caption{Muons traveling through 10 meters of iron. Dashed lines indicate linear regression fits.}
        \label{fig:benchmark}
    \end{subfigure}
    \hfill
    \begin{subfigure}[b]{0.38\textwidth}
        \centering
        \vspace{3pt}
        \begin{tabular}{lc}
            \toprule
            Method & Time (s) \\
            \midrule
            Geant4* & 4274881.6 \\
            Geant4 (1024) & 11279.2 \\
            Our Method & 88.3 \\
            \bottomrule
        \end{tabular}
        \caption{Propagation of 500M muons through the SHiP's Muon Shield. Single core Geant4 time is extrapolated from a linear fit.}
        \label{tab:benchmark}
    \end{subfigure}
    \caption{Timing comparison between Geant4 and our GPU-accelerated method.}
\end{figure}

Finally, we present a timing benchmark comparing the performance of our GPU-accelerated method against the Geant4 simulation. For both experiments presented here, we show results for single-core Geant4 and multi-threaded Geant4. For the Muon Shield specifically, we compared with the results obtained using a CPU cluster with 1024 cores. 
The results are summarized in Table~\ref{tab:benchmark}. Our data-driven GPU approach, running on a NVIDIA L40S, achieves a speedup of more than four orders of magnitude compared to single-core Geant4. In particular, it remains over $100\times$ faster than the Geant4 simulation running on 1024 CPU threads. It is also worth noting that our architecture supports multi-GPU scaling, allowing for further linear reductions in runtime by distributing particle batches across multiple devices.

\section{Discussion and Conclusion}
In this work, we have presented a GPU-accelerated framework for Monte Carlo particle transport based on histogram sampling of precomputed interaction statistics. By separating the computationally intensive physics modeling into an offline phase and performing the online transport entirely on GPUs, our method achieves a dramatic reduction in simulation runtime over traditional softwares. Specifically, benchmarks demonstrate speedups exceeding four orders of magnitude relative to single-core Geant4 simulations and maintain a factor of $\sim 100\times$ advantage over massively parallel implementations running on 1024-core CPU clusters. Importantly, this performance is achieved without compromising physical fidelity in the studied regime, as confirmed by the excellent agreement with Geant4 observables in complex magnetic environments. Moreover, this new approach allows large-scale simulations to be executed on a single GPU-equipped workstation, removing the need for extensive CPU clusters and their associated management overhead.

Our method occupies a middle ground between first-principles analytical simulators, such as Geant4, and fully end-to-end generative models. While Geant4 remains the gold standard for high-fidelity physics, it incurs a high computational cost. Conversely, deep learning approaches often act as ``black boxes'' that replace the entire simulation chain. Our histogram-based approach bridges this gap: it achieves the computational efficiency of generative models while maintaining the interpretability and step-by-step transport logic of a Monte Carlo simulation.

Another compelling advantage of our histogram-based formulation is its potential for differentiable simulation. There is a growing effort within high-energy physics to develop differentiable transport codes, which would enable gradient-based optimization of detector designs and reconstruction algorithms \parencite{adelmann2022new, aehle2023progress}. While making the full Geant4 codebase differentiable is an immense technical challenge, our method simplifies the interaction physics into sampling operations from static probability densities. By applying established techniques for differentiable sampling --- such as the reparameterization trick or Gumbel-Softmax relaxation --- our framework could be extended to propagate gradients through the transport process. This would effectively transform the simulation into a differentiable layer within a larger machine learning pipeline, allowing for the direct, end-to-end optimization of detector geometries and magnetic field configurations via backpropagation.

Although the method was developed for a specific class of problems—those dominated by stochastic momentum loss and multiple scattering—it remains broadly applicable across a range of transport scenarios. In particular, its efficiency makes it well-suited for iterative or high-statistics studies where conventional Monte Carlo simulations would be prohibitively slow. One natural extension is to neutrino or lepton transport through dense media, where importance sampling or biasing schemes could be integrated to handle rare interactions efficiently. Similarly, domains such as cosmic-ray propagation, radiation shielding design, and macroscopic dose estimation in medical physics represent promising applications where high-throughput transport is required and statistical precision is prioritized over microscopic event-by-event detail.

Future work can potentially focus on extending the technique to include additional processes such as particle decay, secondary generation, and adaptive step sizes. Furthermore, while this work focused on particle physics, the underlying methodology is broadly applicable. The technique of reducing complex interactions into precomputed probability distributions can be generalized to any problem governed by similar stochastic transport dynamics, offering a scalable path for acceleration across scientific domains.

\section*{Acknowledgments}
We acknowledge support from the Swiss National Science Foundation through the Advance Grant271
TMAG-2 209263.272.

\clearpage
\printbibliography

\appendix

\section{Optimal Momentum Binning}
\label{appendix:momentum_binning}

As discussed in Section \ref{sec:building_histograms}, the physical processes governing particle transport exhibit a strong non-linear dependence on the particle's momentum $p$. A uniform binning strategy is inefficient because it necessitates an excessively fine granularity to capture the rapid variations at low momenta, resulting in redundant bins in the high-momentum regime where the physics is relatively stable.

Although for our purposes a simple logarithmic binning proved sufficient, different applications may require a more tailored approach. To address this, we define an optimal binning strategy based on the principle of bounded variation. The objective is to partition the momentum spectrum $[p_{min}, p_{max}]$ into a set of $N$ disjoint intervals (bins) $\{ [p_0, p_1), [p_1, p_2), \dots, [p_{N-1}, p_N] \}$, where $p_0 = p_{min}$ and $p_N = p_{max}$. The optimality condition requires that the bin widths are maximized while ensuring that the variation of the physical observables within each bin remains below a specified tolerance threshold $\epsilon$.

Let $\Phi(p)$ denote the physical quantity of interest describing the interaction statistics at momentum $p$. In the absence of a closed-form analytical expression, $\Phi(p)$ must be estimated empirically. This is achieved by performing a "pilot" simulation where, for any given probe momentum $p$, we simulate a batch of $K$ particles and compute the sample mean. For our specific case, the estimator $\hat{\Phi}(p)$ is defined as:
\begin{equation}
    \hat{\Phi}(p) = \frac{1}{K} \sum_{k=1}^{K} \log\left(- \frac{\Delta P_{z,k}(p)}{|\mathbf{P}|}\right)
\end{equation}
where $\Delta P_{z,k}(p)$ is the momentum loss of the $k$-th particle with initial momentum $p$. Alternatively, $\hat{\Phi}(p)$ can be defined as a vector function comprising estimators for both energy loss and scattering width.

We define the variation $\Delta(p_a, p_b)$ over a candidate bin interval $[p_a, p_b]$ as the maximum absolute difference between the estimator at the lower bin edge and the estimator evaluated at any point within the interval:
\begin{equation}
    \Delta(p_a, p_b) = \sup_{p' \in [p_a, p_b]} | \hat{\Phi}(p') - \hat{\Phi}(p_a) |
\end{equation}
In practice, the supremum is approximated by evaluating $\hat{\Phi}(p')$ on a fine discrete grid of probe momenta within $[p_a, p_b]$.

The sequence of optimal bin edges $\{p_i\}$ is then constructed recursively. Given a bin edge $p_i$, the subsequent edge $p_{i+1}$ is defined as the upper bound of the largest contiguous interval starting at $p_i$ that satisfies the tolerance constraint:
\begin{equation}
    p_{i+1} = \sup \{ p \in (p_i, p_{max}] \mid \Delta(p_i, p) \le \epsilon \}
\end{equation}

This recursive definition ensures that the bin density $\rho(p) \propto (\frac{d\hat{\Phi}}{dp})$ adapts dynamically to the underlying physics as revealed by the pilot simulation. Regions where the estimated interaction properties change rapidly (large variation, typically at low momentum) result in narrow bins, while regions of stability result in wide bins. It is important to note that, when using this dynamic binning strategy, a binary search is required during the simulation to identify the appropriate bin for a given momentum, as opposed to the direct indexing possible with uniform or logarithmic binning.

\end{document}